%% file: main.tex
\documentclass[sigconf]{acmart}
\usepackage{booktabs} 
\usepackage{graphicx}
\usepackage{multirow}
\usepackage{amssymb,amsmath,amsfonts,scalerel}
\usepackage{amssymb}
\usepackage{color}
\usepackage{rotating}
\usepackage{booktabs}
\usepackage{balance}
\usepackage{hyperref}
\usepackage[ruled,linesnumbered,vlined]{algorithm2e}
\usepackage{lineno}
\usepackage{subcaption}
\usepackage[textsize=scriptsize,backgroundcolor=green]{todonotes}
\usepackage[inline]{enumitem}
\usepackage{xspace}

\newlist{inlinelist}{enumerate*}{1}
\setlist*[inlinelist,1]{%
  label=(\roman*),
}

\newcommand{\nosemic}{\renewcommand{\@endalgocfline}{\relax}}
\newcommand{\dosemic}{\renewcommand{\@endalgocfline}{\algocf@endline}}
\let\oldnl\nl
\newcommand{\nonl}{\renewcommand{\nl}{\let\nl\oldnl}}

\newcommand{\partitle}[1]{\vspace{2mm}\noindent\textbf{#1}}

\newcommand{\appname}{Omicron\xspace}

\begin{document}

\title[Understanding Mobile Search Task Relevance and User Behaviour in Context]{Understanding Mobile Search Task Relevance \\ and User Behaviour in Context}

\author{Mohammad Aliannejadi}
\affiliation{%
  \institution{Universit{\`a} della Svizzera italiana (USI)}
}
\email{mohammad.alian.nejadi@usi.ch}

\author{Morgan Harvey}
\affiliation{%
  \institution{Northumbria University}
}
\email{morgan.harvey@northumbria.ac.uk}

\author{Luca Costa}
\affiliation{%
  \institution{Universit{\`a} della Svizzera italiana (USI)}
}
\email{luca.costa@usi.ch}

\author{Matthew Pointon}
\affiliation{%
  \institution{Northumbria University}
}
\email{m.pointon@northumbria.ac.uk}

\author{Fabio Crestani}
\affiliation{%
  \institution{Universit{\`a} della Svizzera italiana (USI)}
}
\email{fabio.crestani@usi.ch}

\begin{abstract}
Improvements in mobile technologies have led to a dramatic change in how and when people access and use information, and is having a profound impact on how users address their daily information needs. Smart phones are rapidly becoming our main method of accessing information and are frequently used to perform ``on-the-go'' search tasks. As research into information retrieval continues to evolve, evaluating search behaviour in context is relatively new. Previous research has studied the effects of context through either self-reported diary studies or quantitative log analysis; however, neither approach is able to accurately capture context of use at the time of searching.

In this study, we aim to gain a better understanding of task relevance and search behaviour via a task-based user study (n=31) employing a bespoke Android app. The app allowed us to accurately capture the user's context when completing tasks at different times of the day over the period of a week. Through analysis of the collected data, we gain a better understanding of how using smart phones on the go impacts search behaviour, search performance and task relevance and whether or not the actual context is an important factor.
\end{abstract}


\copyrightyear{2019} 
\acmYear{2019} 
\setcopyright{acmcopyright}
\acmConference[CHIIR '19]{2019 Conference on Human Information Interaction and Retrieval}{March 10--14, 2019}{Glasgow, United Kingdom}
\acmBooktitle{2019 Conference on Human Information Interaction and Retrieval (CHIIR '19), March 10--14, 2019, Glasgow, United Kingdom}
\acmPrice{15.00}
\acmDOI{10.1145/3295750.3298923}
\acmISBN{978-1-4503-6025-8/19/03}

\maketitle

\input{sec-introduction}

\input{sec-related-work}

\input{sec-method}

\input{sec-results}
\input{sec-discussions}
\input{sec-conclusion}

\bibliographystyle{ACM-Reference-Format}
\bibliography{main}

\end{document}

%% file: sec-introduction.tex
\section{Introduction}
\label{sec-introduction}
Recent years have seen the ever-increasing use of mobile devices, such as smartphone and tablets, to search the web for information, to the extent that, as of 2016, more searches are performed on mobile devices than on more ``traditional'' desktop computers~\cite{globalstats:2016}. Mobile devices can be readily used in many situations that a desktop computer cannot and, since they are carried constantly on one's person, at most of the times of the day and night. As such, searching is now performed in a larger range of contexts than ever before and, often, contemporaneously with other tasks~\cite{harvey:2017}.

Context refers to the the circumstances that form the setting for an event (e.g. location, time or current activity) and can have a profound effect on task relevance and on people's information needs, their search behaviour and ultimate success~\cite{hinze:2010,church:2014}. Research shows that mobile users search throughout most of the day~\cite{song:2013} - often from immediately after waking in the morning until falling asleep at night - and in many different locations~\cite{kaikkonen:2008,harvey:2017}, including when ``on the go'' walking down the street, in pubs and restaurants and on public transport.

Previous research has investigated how context affects mobile search task relevance, search behaviour, performance and success typically by means of either self-reported diary studies (e.g.~\cite{church:2014,hinze:2010,komaki:2012}) or through the quantitative analysis of large-scale search logs (e.g.~\cite{kamvar:2006}). While both approaches can provide useful insights to better understand and improve mobile search, neither is able to accurately capture the context of the user at the time of searching. Log studies are unable to easily capture user's opinions and feelings regarding the task and self-reported diary studies rely on the ability of the user to accurately and honestly report on the context and on their own searching behaviour. 

In this work we seek to gain a more complete understanding of task relevance and search behaviour by means of a task-based user study. Our main research questions, therefore, are:
\begin{itemize}
\item{\textbf{RQ1:}} Which kinds of task do people prefer to complete on mobile devices and what impact does time have on this?
\item{\textbf{RQ2:}} How does perceived task relevance vary by temporal context and what impact does this variation have on user behaviour and performance?
\item{\textbf{RQ3:}} What effect does situational context (activity) have on performance?
\end{itemize}

To this end, we have developed a bespoke Android application (aka. app), called \textit{\appname}, that prompts the participants to complete a given search task at a specific time window. The participants were asked to keep \appname running on their phones for a period of one week. During this week, a background service kept collecting useful information about the participants context as well as their interactions with their device, while searching. We have made \appname open source to facilitate research on mobile search\footnote{Available at \url{https://github.com/aliannejadi/Omicron}}.

The remainder of the paper is structured as follows: In section~\ref{sec-related-work} we consider related work on the topics of query log analysis, mobile search interaction and interaction with mobile devices in general; section~\ref{sec-method} describes our research methodology including the user studies we performed; sections~\ref{sec-results} describes the results of the user studies in detail; section~\ref{sec-discussions}
discusses how the results relate to the existing literature and suggests reasons and intuition behind them; and section~\ref{sec-conclusion} concludes the paper with suggestions for potential future work.

%% file: sec-related-work.tex
\section{Related Work}
\label{sec-related-work}
Research on human interaction with mobile devices has been growing in recent years. One of the main focuses in this field of research has been to understand how users interact with search engines on mobile devices~\cite{DBLP:series/sbcs/CrestaniMS17}. In this section, we start with brief overview of studies that analyse mobile query logs. We then consider those that analyse human interaction when performing mobile search and, finally, we summarise studies that try to understand human interaction with mobile devices thorough various methods of study.

\subsection{Query log analysis} 
Studying large-scale query logs gives researchers crucial information on how people translate their information needs in the form of queries.
\citet{kamvar:2006} conducted one of the earliest studies, analysing a large-scale mobile search query log and found that mobile queries were less diverse than desktop queries and that the distribution over task types varied between desktop and mobile search.
\citet{DBLP:journals/jasis/CrestaniD06} compared spoken and written queries in a lab study with 12 users. They found that spoken queries are longer and closer to natural language. 
\citet{DBLP:conf/mhci/ChurchSBC08} analysed six million search queries in a period of one week to understand the click-through rate and found that users generally focused only on the first few search results.
Later, \citet{song:2013} studied a commercial search log and found significant difference in search patterns dependent on the devices used (between iPhone, iPad, and desktop). For instance, they found that query length on mobile devices were longer. However, they suggested that the query length continues to change and this could be a sign of evolving mobile usage patterns. Also, query categories, usage time, and location of usage were different among different devices.
More recently, \citet{DBLP:conf/sigir/Guy16} analysed 500,000 spoken queries from a commercial mobile search app, submitted via a voice interface. The analyses showed that voice queries are longer on average and are closer to natural language. Moreover, they are more focused on multimedia content and require less interaction with the device's touchscreen.

\subsection{Mobile search interaction}
Understanding human interaction while doing mobile search has become an area of interest since mobile devices are constantly evolving \cite{song:2013}. For this reason, many researchers have conducted user studies to understand various aspects of user behaviour and interaction in relation to mobile search. 
\citet{DBLP:conf/chi/SohnLGH08} conducted a two-week diary study from 20 participants in which they found that contextual features such as activity and time influence 72\% of mobile information needs.
\citet{kaikkonen:2008} asked 390 mobile Internet users to fill an online survey, followed by 23 face-to-face interviews and analysed the impact of location and Web page design on mobile phone browsing behaviour. Similar to the results of~\cite{kamvar:2006}, the author found that there was a considerable difference between the tasks that mobile and desktop users conducted and that location was an important factor in determining what people wanted to search for using their mobile devices.

\citet{DBLP:conf/iui/ChurchS09} studied the intent behind mobile information needs of twenty users over four weeks via a diary study. They observed significant differences between mobile and desktop information needs. In particular, they found that users had many non-informational information needs, with geographical and personal information needs being popular.
Later, \citet{DBLP:conf/mhci/ChurchO11} carried out another diary and interview study to understand the shift of mobile information needs at the time of the study. The study was done over a four-week period with 18 active mobile users, discovering that the popularity of stationary mobile Web access was increasing. 
In another attempt to understand users' information needs, \citet{church:2014} conducted a large-scale \textit{snippet-based diary} \cite{DBLP:conf/chi/BrandtWK07} study with 100 participants throughout a three-month period. This technique allowed users to capture moments in-situ and send them via SMS or MMS. Later, they could access a Web site in which they would review the messages and provide more details about their context. They found significant differences in terms of information needs and how they were addressed depending on user gender, device and location.

Other studies have aimed to analyse touch-screen gestures \cite{DBLP:conf/www/WilliamsKCZAK16}, the effect of searching on-the-go \cite{harvey:2017}, the effect of result snippets \cite{DBLP:conf/chiir/KimTSGY17} as well as users' perception of result usefulness~\cite{DBLP:conf/ecir/MaoLKL0M18}.
\citet{DBLP:conf/www/WilliamsKCZAK16} conducted a lab study with 60 participants, focusing on the analysis of user gesture interactions, such as touch actions, and their relation with good search abandonment on mobile search. They showed that the time spent interacting with answers on a SERP is positively correlated with good abandonment and satisfaction.
Through another lab study with 72 participants, \citet{DBLP:conf/sigir/OngJSS17} observed different patterns in user behaviour while doing mobile and desktop search as the amount of information scent was altered. They found that users' behaviour differ significantly on mobile: for instance, desktop users preferred SERPs with a higher number of relevant search results; whereas this preference was not observed in the mobile environment.
\citet{DBLP:conf/chiir/KimTSGY17} conducted a lab study with 24 participants and analysed the effect of snippet size on mobile search time and accuracy. They found that, for informational tasks on mobile devices, longer snippets lead to longer search times with no better search accuracy.

\subsection{Situational context}

Since mobile devices can be used in many different contexts (e.g. when walking on when on public transport), researchers have considered what impact these situations have on users as they go about various tasks.
\citet{DBLP:conf/chi/BragdonNLH11} conducted a lab study with 15 participants and found that, in the presence of environmental distractions, touch-screen gesture design leads to significant performance gain and reduced cognitive load, as opposed to soft buttons. In another lab study of 13 participants, \citet{DBLP:conf/mhci/MizobuchiCN05} analysed the speed and accuracy of text input on mobile devices under various situational contexts. They found that texting whilst walking resulted in either a significant decrease in input speed or walking pace, indicating that the mental workload of performing both tasks simultaneously results in the user prioritising one over the other.

\citet{barnard:2007} investigated reading comprehension and word search when walking and found that contextual variations can have large effects on user behaviour by impairing performance and increasing user workload. \citet{harvey:2017} recruited 24 participants and conducted a lab study where they found that fragmented attention of users while searching on-the-go affects their search objective and their perception of task difficult and of their own performance.
\citet{hoggan:2009} considered the effect of using public transport while performing simple touchscreen
typing tasks. Participants performed the tasks on a noisy and
bumpy subway and were found to perform progressively worse as
the noise level and bumpiness of the ride increased. \citet{harvey:2018} confirmed this finding in a lab study by simulating noisy everyday situations. In this work participants were asked to complete search tasks and those subjected to the noisy conditions were found to perceive more stress and time pressure, leading to a reduced ability to identify task-relevant documents and a compulsion to finish the search task more quickly.

\subsection{Limitations of prior research} 
Although previous research has studied mobile search and the effect of context in various aspects, it has been done via lab-based, self-reporting and/or diary studies. Each of these study settings introduce specific limitations. In particular, a lab study does not capture users' interactions in-situ, as users' attention varies dramatically depending on their context. Moreover, self-report and diary studies highly depend on participants' willingness to report their context and details of their searches and on their ability to accurately recall and describe their interaction and behaviour. Therefore, many contextual information would be missing in situations when users are distracted. In this study, we perform a task-based field study, aiming to understand the impact of context on users' behaviour in-situ. This enables us to examine users' willingness to complete various mobile search tasks as well as their engagement, interaction, behaviour and performance under various contexts.

%% file: sec-method.tex
\section{Method}
\label{sec-method}
To achieve our aims we conducted a task-based user study in which participants were asked to complete a number of search tasks based on a pre-defined schedule over the period of a week (from Monday through to Sunday). Each participant was randomly assigned to one of three ``groups'', each of which had a different schedule of times and days when they would be issued with tasks. Table~\ref{tab:schedule} shows an example group schedule. The schedules were designed such that, over the groups, tasks were evenly balanced over all time slots and such that users received a maximum of two task notifications per weekday and only one on Saturday and Sunday. This was to ensure that participants were not overloaded and would thus be more likely to complete all of the tasks when scheduled. The 5 time periods were intended to capture different temporal contexts during the user's day, for example the ``Early morning'' period (between 8AM and 10AM) should capture the context of the user either preparing to go to work, walking to work or just starting work in the morning.

\begin{table}[t]
\caption{Example group task schedule.}
\label{tab:schedule}
\begin{tabular}{l|lllllll}
                & Mon & Tue & Wed & Thur & Fri & Sat & Sun \\ \hline
Early morning   & x   &     &     & x    &     &     &     \\
Late morning    &     &     & x   &      & x   &     & x   \\
Early afternoon & x   & x   &     & x    &     &     &     \\
Late afternoon  &     &     & x   &      & x   &     &     \\
Evening         &     & x   &     &      &     & x   &    
\end{tabular}
\end{table}

Participants were required to install \appname on their mobile phones for the duration of the study. The app alerted users to new tasks, allowed them to complete tasks by searching and indicating relevant documents, presented users with questionnaires and logged interactions and various other sensor data. Data collected included: acceleration and gyroscopic sensor data (which could be used to infer activity), battery level, GPS location and light level. Activities included in vehicle, walking, on bicycle, running and still and were inferred based on accelerometer and gyroscope data by Google's Activity Recognition Transition API~\footnote{https://developer.android.com/guide/topics/location/transitions}. All collected data was automatically uploaded to a secure server for analysis; a data recovery system was implemented to ensure that all recorded data reached the remote server, even if there were times when no connection was available.

The 5 search tasks were designed to conform to the kinds of tasks people have been shown to frequently conduct on mobile devices~\cite{song:2013,carrascal:2015}. Some task types are further delineated into individual ``one-shot'' tasks (i.e. those that a single user could only be given once) and ``multi-use'' tasks. The tasks were as follows:

\begin{enumerate}
\item  Entertainment (multi-use) \\
You are feeling a bit bored. Find a film you would like to watch.
\item  Restaurant/local (multi-use) \\
You are feeling hungry. Find a restaurant you want to go to to eat.
\item Shopping/cooking (multi-use)
Find a recipe you'd like to cook for your dinner tonight.
\item Information look-up (single-use)
    \begin{enumerate}
    \item Find out what the latest headlines are in {\em XX}.
    \item You are looking for a new car. Find out what electric vehicles are available in your country.
    \item You are soon off on holiday to {\em XX}. Find some fun/interesting things to do there.
    \end{enumerate}
\item Transport (single-use)
	\begin{enumerate}
    \item Find a cheap way to get to {\em XX} from {\em YY}.
    \item You fancy some exercise. Find a nice walk from where you are to {\em XX}.
    \item You are planning a bike ride around the city. Where can you hire a bike?
    \end{enumerate}
\end{enumerate}

\begin{figure}[h]
\centering
\begin{subfigure}{.25\textwidth}
  \centering
  \includegraphics[width=.9\linewidth]{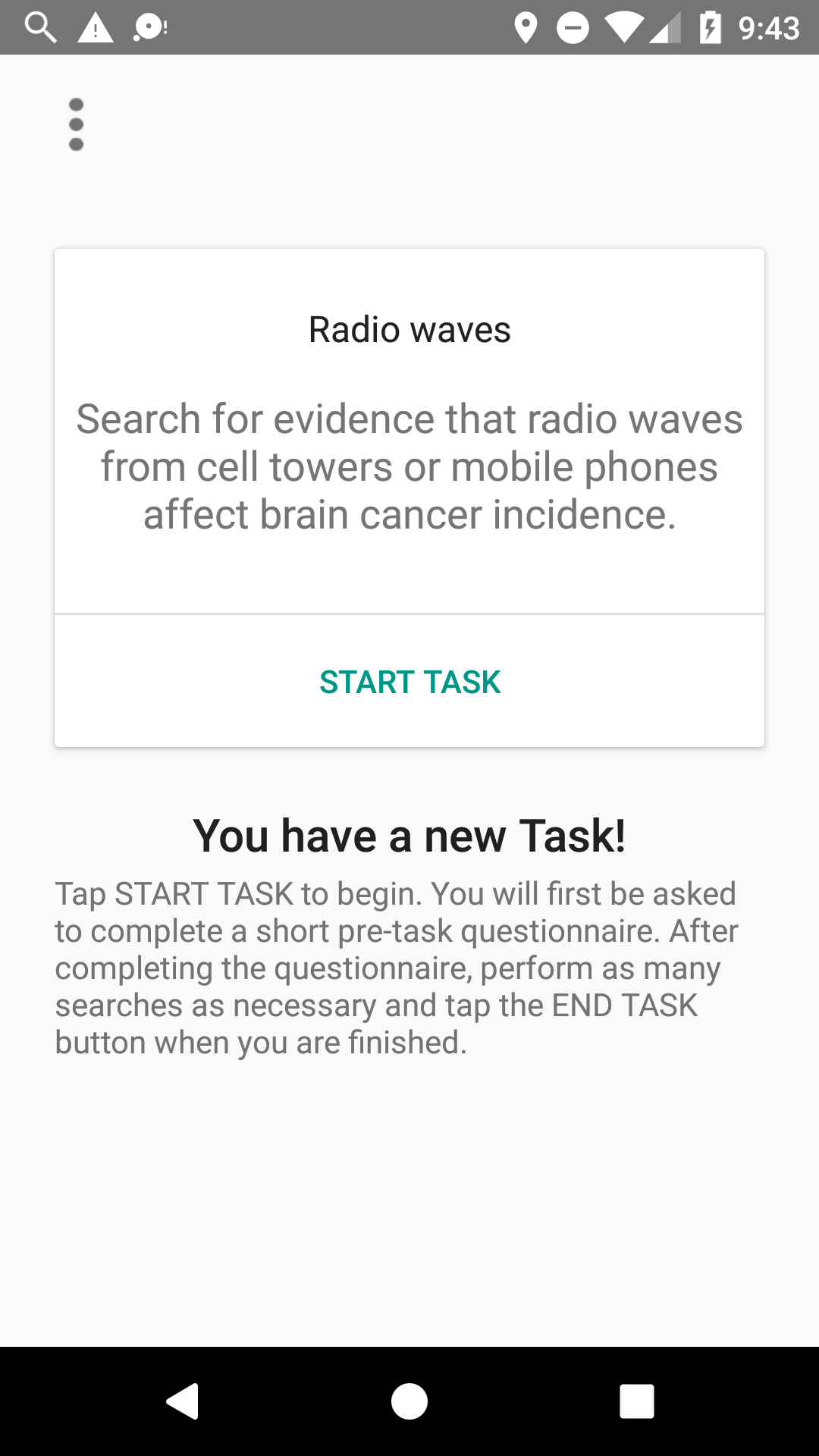}
  \caption{New task notification.}
  \label{fig:new_task}
\end{subfigure}%
\begin{subfigure}{.25\textwidth}
  \centering
  \includegraphics[width=.9\linewidth]{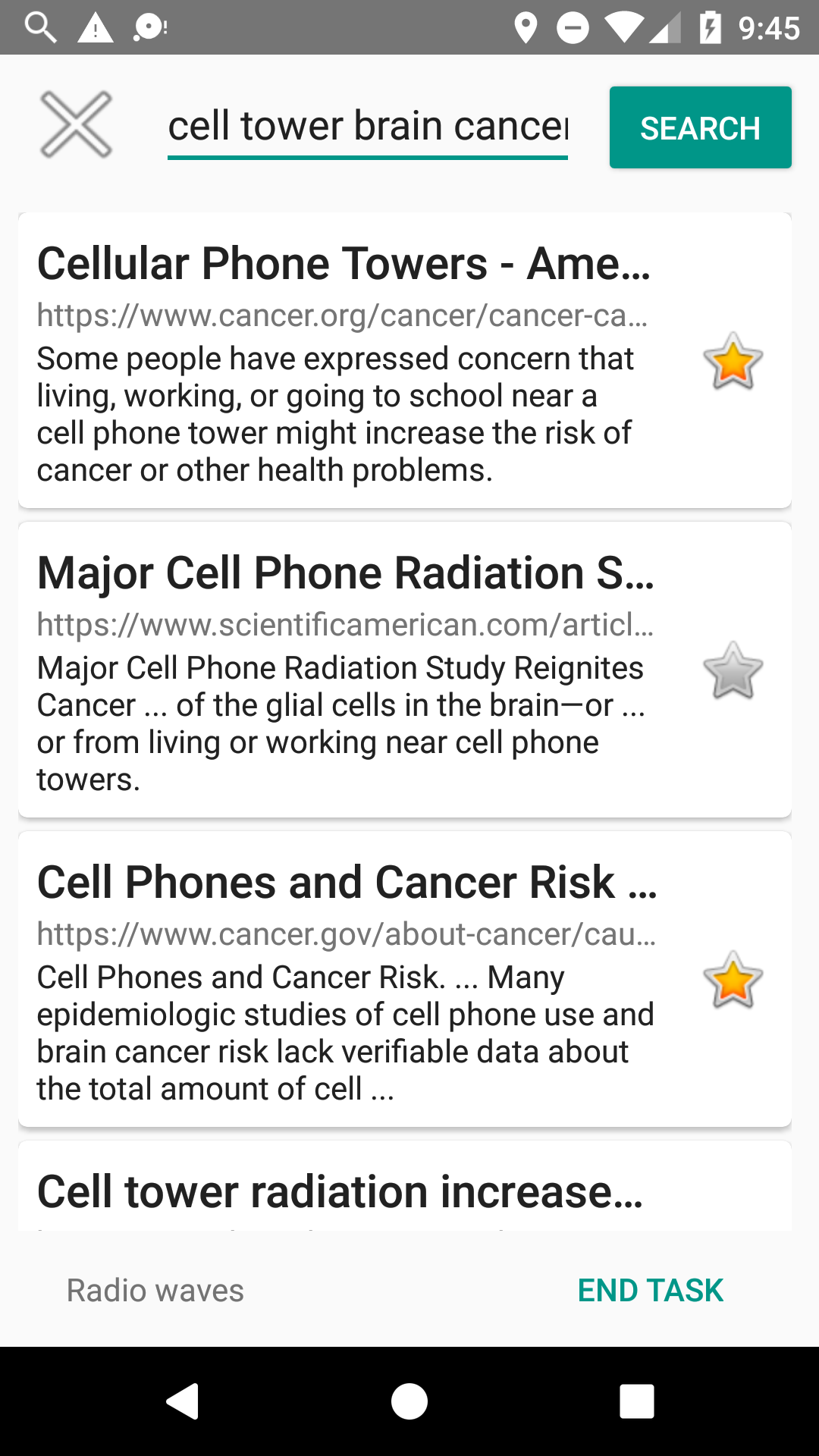}
  \caption{App search interface.}
  \label{fig:search_screen}
\end{subfigure}
\caption{Interface screenshots.}
  \label{fig:screenshots}
\end{figure}

Tasks were automatically sent to participants' phones based on their own group-based schedules and in a random order (see Figure~\ref{fig:new_task}). Once a new task was received by the device, a system prompt would appear and persist in the system notification tray until either actioned or dismissed. Participants could choose to dismiss a task, in which case it would be scheduled to reappear at a later time slot, unless there was another task scheduled for that period, in which case the delayed task would appear directly after the new task. Once assigned a new task, participants were able to use the app's own search interface to search for, read and bookmark relevant documents (see Figure~\ref{fig:search_screen}). Search was implemented by using the Bing search API and displayed short snippets for each result. Participants could indicate relevance by tapping star icons next to each search result. Note that participants were given a test task after installing the app and were given a tutorial explaining the interface.

Before and after completing each task participants were asked to complete short questionnaires, which appeared automatically within \appname and were administered by Qualtrics\footnote{\url{https://www.qualtrics.com}}. The pre-task questionnaire asked participants about their prior knowledge of the given task, how difficult they expected it to be, whether or no the task was clear and their level of interest in the task. After completing the task they were asked about how stressed and distracted they were feeling, how difficult they found the task, how relevant the task was the them personally, how relevant it was to the current context (time of day) and how often they perform such a task. Additionally, after installing the app on their device, participants completed a short pre-study questionnaire about their demographic details and mobile device usage.

%% file: sec-results.tex
\section{Results}
\label{sec-results}
\subsection{Participants}
Participants were recruited by researchers from both collaborating institutions in Switzerland and the UK. In total 31 people took part in the study, of whom 21 were male and who had a median age range of 25 to 34 years (only 3 participants were younger than this and only 2 older). Participants were generally well-educated - 23\% held Bachelor's degrees, 58\% Master's and 6\% doctorates - and 74\% described their English language ability as advanced or better; there were 6 native speakers.
14 participants stated that they most often use their mobile phones to connect to the Internet, 11 use a laptop and 6 most often use a desktop machine.  13 reported using their mobile device between 1 and 2 hours per day on average, 9 used theirs between 2 and 3 hours, and 8 said they use their phone for more than 3 hours per day; only a single participant reported using their device for less than an hour per day.

\subsection{Task completion}

Each participant was assigned a total of 12 tasks to complete over the week: 2 per weekday and 1 each on Saturday and Sunday. Of the 31 participants, only 5 completed all 12 tasks, with a median completion rate of 9 tasks per participant. Out of a total of 372 assigned tasks, participants rescheduled 123.

\begin{figure} [ht]
\centering
\includegraphics[width=0.48\textwidth]{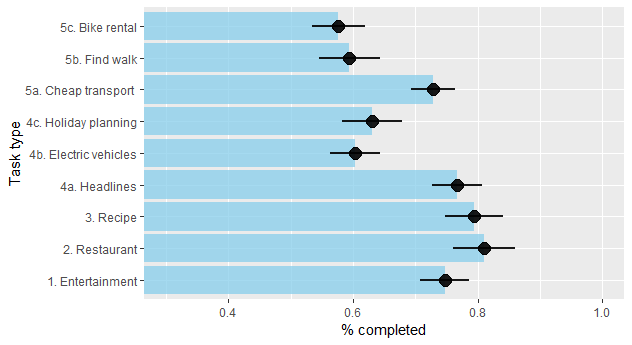}
\caption{Completion rate by task.}
\label{fig:completion_by_task} 
\end{figure}

Completion rates varied considerably by task type. Figure~\ref{fig:completion_by_task} shows the completion rate by type of task (and 95\% confidence interval\footnote{Estimated via empirical analysis of a Bootstrap sampling with 1,000 resamples.\label{footnote:bootstrap}}; please refer to Section~\ref{sec-method} for expanded task descriptions). Although linear modelling did not demonstrate task to be a significant predictor of completion rate, it does appear that the task had some effect. Participants seemed to be more willing to complete tasks relating to food, eating and news headlines (tasks 2, 3 and 4a) than on bike rental and electric cars. This may be because cuisine and news are topics relevant to almost everyone, while cycling and automotive subjects may only be relevant to a smaller number of people. It may also be that people perceived the food-related tasks to be easier to complete or more interesting.

Analysis of the pre-task questionnaire data confirms some of these inferences. There was a moderate positive correlation between average completion rate and the mean response to the question of how knowledgeable participants were about the task ($\rho(9) = 0.536, p = 0.137$), a strongly (and significant) negative correlation with how difficult participants expected the task to be ($\rho(9) = -0.821, p = 0.007$; the more difficult the task was perceived to be, the less likely it was that it would be completed), and a moderate positive correlation between completion rate personal interest in the task ($\rho(9) = 0.379, p = 0.314$).

\begin{figure} [h]
\centering
\includegraphics[width=0.48\textwidth]{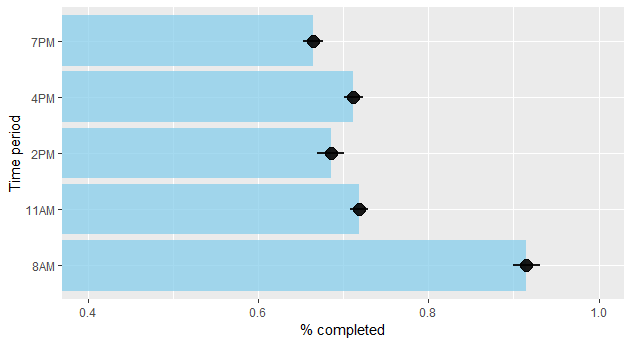}
\caption{Completion rate by scheduled time period.}
\label{fig:completion_by_time_period} 
\end{figure}

Completion rates also varied quite considerably depending on the time period in which the task was given to the user. Figure~\ref{fig:completion_by_time_period} shows the completion rates and 95\% confidence intervals\textsuperscript{\ref{footnote:bootstrap}} over all participants by time period. This demonstrates that users were much more likely to accept and complete a task in the morning but that this willingness to engage reduced as the day went on with almost half of all tasks assigned at 7PM being ignored. Logistic regression modelling shows that the time period is a significant predictor of completion ($\beta = -.11, t(372)=-3.65, p \ll 0.01$).

\begin{figure} [h]
\centering
\includegraphics[width=0.48\textwidth]{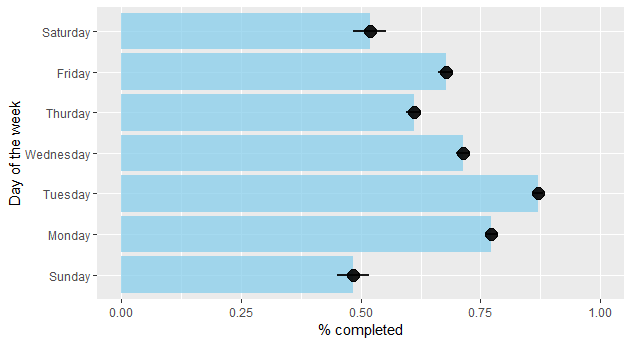}
\caption{Completion rate by weekday.}
\label{fig:completion_by_weekday} 
\end{figure}

As shown in Figure~\ref{fig:completion_by_weekday}, the weekday on which the task was assigned has a large impact on completion rate. Perhaps unsurprisingly, the completion rate is worst on weekends, very good on the first couple of days of the study (i.e. Monday and Tuesday) and middling in later weekdays. The weekday is also a significant predictor of completion rate ($\beta = -.1, t(372) = -1.87, p = 0.031$).

\subsection{Pre-task perception}

\begin{figure} [h]
\centering
\includegraphics[width=0.48\textwidth]{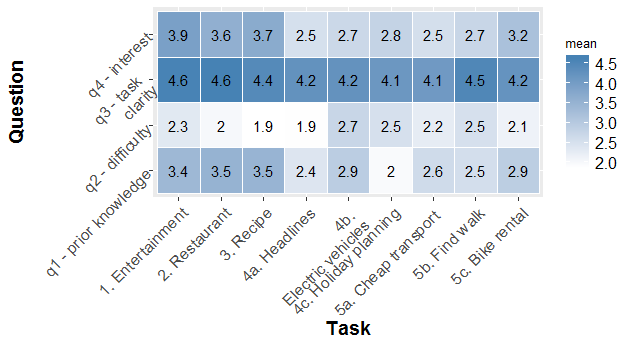}
\caption{Mean response to pre-task questionnaire by task type.}
\label{fig:task_pre_heatmap} 
\end{figure}

Before completing each task, participants filled in a set of four pre-task questions to gauge their prior knowledge of, and interest in the topic, how difficult they expected the task to be and how clear they felt the task was. Responses to the questions on prior knowledge and difficulty varied little by either weekday or allocation time, although there was a trend of expected difficulty monotonically increasing from a low on Mondays (mean = 1.92) through to a high on Saturdays (mean=2.43). Linear modelling showed the weekday to be a significant predictor of expected difficulty ($\beta = 0.09, t(526) = 3.59, p \ll 0.01$), perhaps suggesting that the cumulative impact of work over the week can make people feel less able to complete the tasks well. Neither task clarity nor interest showed any noticeable patterns by either weekday or start time.

Figure~\ref{fig:task_pre_heatmap} shows how responses to these questions varied by task type. It seems that participants found all of the tasks to be clear and straightforward, with only very small variations in reported task clarity (Q3). Prior knowledge (Q1) varied quite considerably: participants were quite knowledgeable about the cuisine- and entertainment-based tasks but reported knowing less about news headlines and planning a trip to Slovenia. There was some variation in task interest, which has a similar pattern to that of prior knowledge (i.e. users tend to prefer tasks they feel they are knowledgeable about); responses to these two questions were significantly positively correlated ($\rho(526) = 0.499, p \ll 0.01$).

Difficulty (Q2) was generally assessed to be quite low, although the questions on electric vehicles and holiday planning were expected to be most difficult. We can compare the expected difficulty of tasks from the pre-task questionnaire with the responses to the same question on difficulty in the post-task questionnaire. In doing so we find that most tasks were actually easier than participants expected, those tasks that were expected to be most difficult (4b and 4c). The only exception was task 5c, which participants rated as being more difficult to complete than they had expected. 

\subsection{Task relevance and post-task perception}

\begin{figure} [h]
\centering
\includegraphics[width=0.48\textwidth]{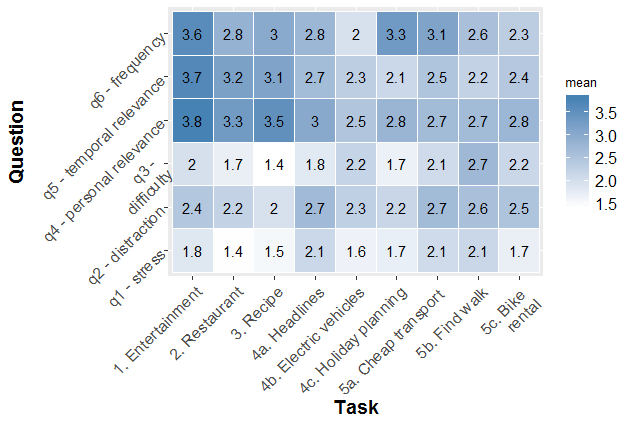}
\caption{Mean response to post-task questionnaire by task type.}
\label{fig:task_post_heatmap} 
\end{figure}

Figure~\ref{fig:task_post_heatmap} shows the mean post-task questionnaire responses by task. Participants did not generally find the tasks to be too difficult, with the possible exception of {\em question 5b.} Interestingly, the level of perceived stress and distraction does seem to have been somewhat dependent on the task, with the transport-related tasks generally producing more feelings of stress. The first 3 tasks - on entertainment and eating - were deemed to be the most personally relevant on average, while the question on electric cars was generally the least personally relevant. We further investigate temporal relevance in the next subsection. 

Interestingly, question 4 (interest in topic) from the pre-task questionnaire is a significant predictor of questions 4, 5 and 6 in the post-task questionnaire (all p-values $\ll 0.001$). This indicates that pre-task interest and temporal and personal relevance are strongly related and that those tasks that a user performs frequently are judged to be more interesting. 

\begin{figure} [h]
\centering
\includegraphics[width=0.48\textwidth]{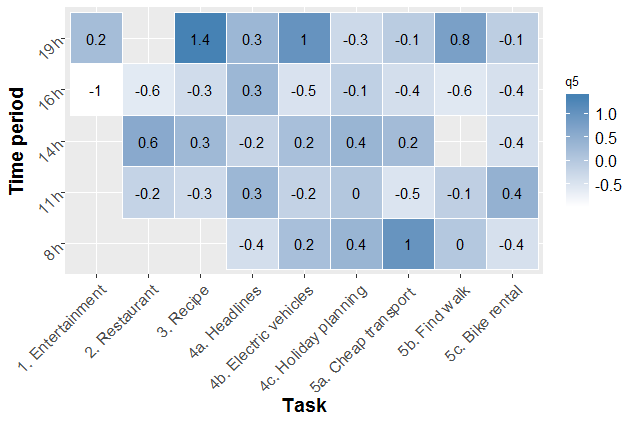}
\caption{Temporal relevance by scheduled time period (normalised by the mean response for each task).}
\label{fig:relevance_by_time} 
\end{figure}

It is interesting to consider how the combination of task and time period affects participants' perceptions of task relevance. As part of the post-task questionnaire, we asked participants to rate the temporal relevance of the task (original question: ``How relevant did you find this task to your current context (time
of day)?'') from 1 - extremely irrelevant to 5 - extremely relevant. We also asked participants to comment on their rating. Figure~\ref{fig:relevance_by_time} shows the mean response by task type and time period (note that task and time period combinations with fewer than 3 responses are left blank and that the responses have been normalised by subtracting the mean response for each task). 

The entertainment task appears to be more relevant in the evening than in the mid-afternoon (where participants found it to be very non-relevant), perhaps because people are at work in the mid-afternoon and, as such, are not able to relax by watching a film, while in the evening they can. The food tasks are thought to be more relevant when they are assigned nearer to the times that people generally eat. This is particularly clear from the recipe task, which is much more relevant in the evening and early afternoon time periods than during other times of the day. The transport finding task is much more relevant in the early morning than it is later in the day, perhaps because people are considering how they will get to work.

If a participant chose the lowest response for the question on contextual relevant (i.e. ``extremely irrelevant'') we asked them to explain why using a textbox. This happened for 32 tasks. In many of these instances the participants explained that they were at work or were busy with other tasks, for example: ``I am busy right now and ate my lunch at 3:30 want focus on my work'', ``I am at work and i [sic] usually check this [sic] things in the evening'' and ``I'm still at work so searching for holiday activities is a distraction''. On 6 occasions participants indicated a lack of relevance of the cycling/bike-related tasks due to a lack of interest, for example: ``Because i [sic] almost never ride bikes'', ``I have no intention of going for a bike ride! Ever!''; or because cycling was not relevant to them at that time: ``I'm having lunch, don't care about biking now'' and ``I'm about to drive into work so don't have time to go fir [sic] a bike ride...''. Users also sometimes showed a lack of interest in the topic of electric cars: ``I'm not buying an electric car now '', ``I don't want to buy an electric car, and it's not a task related to the time''.

\subsection{Searching behaviour and performance}

As we allowed participants to search on the open web, we do not have any relevance judgements available for the tasks we set them. However, we can look at a number of proxy measurements to search performance, as well as indicators of querying behaviour, such as the length of queries, number of queries issued, documents read and number of starred (bookmarked) documents.

\begin{figure} [h]
\centering
\includegraphics[width=0.48\textwidth]{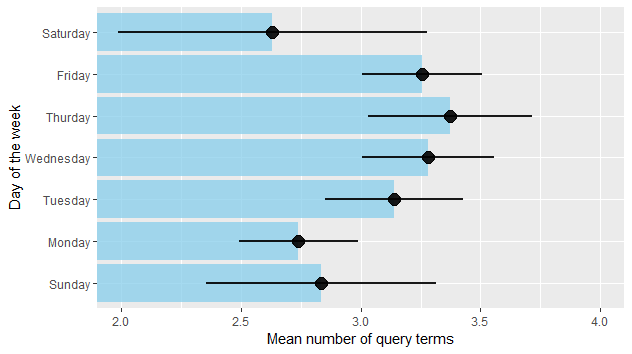}
\caption{Average (mean) query length in terms by weekday.}
\label{fig:ql_by_weekday} 
\end{figure}

The queries submitted were generally quite long at an average length of 3.09 terms and 19.91 characters. Perhaps surprisingly, participants submitted queries of approximately the same length, regardless of the time of day, varying from an average of 2.9 terms in the late afternoon (between 4 and 7PM) to 3.4 in the late morning (between 11AM and 1PM). People tended to spend less time reading search results in the early morning and afternoon (median = 25.5 and 27s) than they did during the rest of the day (median between 31.5 and 36.5s).
There was a much larger variation in query length by day of the week. As shown in Figure~\ref{fig:ql_by_weekday}, it seems that (with the exception of Monday) queries submitted during the week were considerably longer than those submitted over the weekend. We observed a similar temporal effect on result reading time - participants tended to spend more time reading documents on weekdays than during the weekend.

\begin{table}[t]
\caption{Various querying statistics on a per-task basis (q=query; t=task).}
\label{tab:query_stats}
\begin{tabular}{l|llll}
Task & \rotatebox{60}{Terms/q} & \rotatebox{60}{Reading time} & \rotatebox{60}{Bookmarks/t} & \rotatebox{60}{q/t} \\
\hline
1. Entertainment        & 2.68   & 46    & 3.95  & 2.55         \\
2. Restaurant           & 3.05  & 36    & 2.55  & 2         \\
3. Recipe               & 2.92  & 38  & 2.64  & 1.52         \\
4a. Headlines           & 2.59  & 25    & 3.48  & 1.86         \\
4b. Electric vehicles   & 3.24  & 28    & 3.5   & 2.83         \\
4c. Holiday planning    & 3.07  & 29.5    & 3.17  & 1.67         \\
5a. Cheap transport     & 3.31  & 23  & 2.83  & 2.67         \\
5b. Find walk           & 3.89  & 39.5    & 1.59  & 2.24         \\
5c. Bike rental         & 3.14   & 33    & 3  & 2.53        
\end{tabular}
\end{table}

Table~\ref{tab:query_stats} shows querying statistics aggregated by task type. Participants tended to submit much longer queries for certain tasks than others: those submitted to tasks 5b and 5c were significantly longer than those submitted to tasks 1, 3 or 4a~\footnote{Based on Tukey multiple comparisons of means.}. Reading time was broadly similar across tasks, although it was highest for question 1, perhaps because people spent more time reading the synopses of films to decide which they found interesting. Reading time was very low for task 5a, suggesting that, once a suitable query had been generated, it was quite easy to identify relevant documents.

It seems that the tasks deemed to be easiest in the post-task questionnaires (tasks 2, 3 and 4c) required the least amount of query modification, with all 3 of them requiring fewer than 2 queries on average to be completed. Some of the more difficult tasks, on the other hand (e.g. tasks 4b and 5c) required considerably more queries - more than 2 and a half on average. There was also quite a lot of variation in the average number of bookmarks submitted by task. Task 5b had the lowest average (only 1.59), while task 1 had the highest (3.95); there was a significant difference between these two (p$ = 0.03$). Perhaps many participants only felt they needed to find a single walk, while for the film choosing task, despite the fact that the task was to find ``a film you would like to watch'', they wanted to make sure they had alternatives. It may also be because the participants found task 5b to be one of the least interesting and also one of the most difficult.

\begin{table}[t]
\caption{Various querying statistics by activity type (q=query; t=task). * indicates a significant difference compared to the {\it At rest} condition.}
\label{tab:query_stats_by_activity}
\begin{tabular}{l|lllll}
Task & \rotatebox{60}{Instances} & \rotatebox{60}{Terms/q} & \rotatebox{60}{Reading time} & \rotatebox{60}{Bookmarks/q} & \rotatebox{60}{q/t}  \\
\hline
At rest             & 186 & 3.1 & 74   & 1.46 & 2.03  \\
Walking             & 27 & 2.3*  & 48*  & 1.63 & 1.9   \\
After walking & 11 & 3.27 & 50    & 2.55* & 1.54 \\
On transport        & 28 & 2.71 & 68  & 0.7* & 3.11                   
\end{tabular}
\end{table}

\subsection{The effect of situational context}

A key advantage to conducting the study through a bespoke Android app and in a naturalistic setting is that we are able to gain insights into how search behaviour and performance is affected by context and, particularly, the situation in which they perform the search. We logged data from each device's accelerometers and gyroscopes and, using Google's Activity Recognition API, were able to infer from these the activity the user was performing when completing the tasks. Note that in doing so we record the closest recognised activity prior to the submission of queries as being the activity state of the user on query submission. In a small number of cases this state was recorded a considerable amount of time before the queries were submitted and, as such, may be erroneous. To account for this, we only calculate statistics for instances where the queries were submitted within an hour of the closest prior activity.

Table~\ref{tab:query_stats_by_activity} shows the activities inferred by the API (walking, on transport, at rest and after walking) and statistics of the querying behaviour exhibited by participants in these situations. There appears to be considerable difference in a number of querying behaviour metrics depending on the condition. When walking, participants submitted significantly shorter queries than when at rest ($p \ll 0.01$) and spent considerably less time reading search results. These results agree with those of~\citet{harvey:2017}, who observed similar differences in their more artificial, lab-based study and suggest similar causes - i.e. the distractions and attention shifts caused by walking have a deleterious effect on querying behaviour.

When on transport - a difficult condition to accurately simulate in a laboratory - users tended to submit shorter queries, bookmarked significantly fewer relevant documents per query, although reading time was similar to the {\em at rest} condition and they submitted more queries per topic on average. These shorter queries, the tendency to bookmark fewer documents per query and increase in number of queries submitted may be due to the noisy and bumpy conditions experienced when travelling by vehicle. Previous work~\cite{harvey:2018,hoggan:2009} has shown that such conditions can induce a feeling of time pressure and reduce a user's ability to judge the relevance of documents. The increase in number of queries suggests that the users in this condition may be seeking to explore the document space more thoroughly, rather than deeply investigating the documents on each result page.

%% file: sec-discussions.tex
\section{Discussion}
\label{sec-discussions}

This work aims to answer three main research questions in relation to the impact of context on user behaviour and task relevance. In this section, we address each of the research questions in turn.

\partitle{RQ1: Which kinds of task do people prefer to complete on mobile devices and what impact does time have on this?} 
Our results suggest that users prefer to complete tasks on their mobile devices based primarily on two factors: 
\begin{inlinelist}
\item tasks on which they have more prior knowledge and/or interest;
\item tasks that are more relevant to their current context.
\end{inlinelist}
We do not expect users to look for absolutely new information while on the go, unless it is highly relevant to their context. For instance, assume a scenario where a user is having lunch with colleagues and they happen to discuss about the effect of cell towers on brain cancer. The user, presumably, has no prior knowledge about this topic. However, given that he/she is seated (situational context) and seeks to find the answer for the colleagues, it is highly probable that the user completes such a difficult task. Users do not tend to complete search tasks that they perceive to be more difficult, suggesting that they need more attention and time; probably preferring to complete such a task later, either on a desktop or in another context~\cite{kaikkonen:2008}, perhaps where they can better concentrate on it~\cite{DBLP:conf/chi/SohnLGH08}. 
This supports the results of previous work, which showed that users are more engaged with the tasks in which they have more prior knowledge and interest~\cite{DBLP:conf/sigir/Edwards017}.

Our results show that users behave differently at different times of the day as well on different days of the week. In particular, users completed more tasks in the early hours of a day (8AM-11AM) and were typically much less engaged during weekends, similar to the findings of ~\citet{DBLP:conf/mhci/PielotCO14}. It is likely easier for users to engage in a search while their minds are fresh in the morning and less so in the evenings, when other tasks may take priority~\cite{DBLP:conf/chi/SohnLGH08}. 

\partitle{RQ2: How does perceived task relevance vary by temporal context and what impact does this variation have on user behaviour and performance?}
For most of the tasks, the perceived temporal relevance varied based on the time of day. This is more evident for time-dependent tasks such as finding a dinner recipe, which may also be affected by others contextual factors, such as location~\cite{DBLP:conf/chi/SohnLGH08}. Although intuitive, this result is interesting when studied together with search performance, which also varied over the different time periods. Users tended to submit longer queries during weekdays and perform more search tasks in the morning, a result which contrasts with the results of previous research (e.g. \citet{DBLP:conf/sigir/Guy16}). The result also suggest that, while some tasks have more general temporal relevance and, as such, do not vary much in perceived relevance over time, others exhibit much more time-constrained relevance. Such tight constraints on temporal relevance and interest are not generally found in analyses of desktop search log, where querying frequency over the day was much more uniform~\cite{beitzel:2007}. Perhaps in a mobile setting, where users require more attention to understand and complete a given task, they exhibit more variation in their behaviour and are more acutely aware of temporal relevance.

Our results provide useful insights to inform the design of future mobile search features considering both temporal and situational contexts. More specifically, a system should be able to provide query suggestion based on users' past queries at the same time or at the same situation. For instance, assume a user commutes every day on train and needs to check the train schedule and delays on a daily basis. A mobile search system should be able to identify this regular pattern and be able to provide the user with suggested queries~\cite{DBLP:conf/ecir/Bahrainian19} or information cards~\cite{DBLP:conf/sigir/ShokouhiG15,DBLP:conf/chiir/BahrainianC18}.

\partitle{RQ3: What effect does situational context (activity) have on performance?}
We observed substantial variation in behaviour and search performance by situational contexts. Users completed most of the tasks while they were at rest, suggesting that they dismissed the tasks mostly when they were not at rest and waiting until their situational context was more amenable to searching. According to prior research, using mobile devices while walking requires both cognitive and motor abilities~\cite{DBLP:conf/mhci/KaneWS08}; and thus users' attention is fragmented~\cite{harvey:2017}. For this reason, we observed less task engagement, a decrease in participant attention, reduced reading time and a significant reduction in query length whilst walking. 

Our results suggest similar behaviour when users are on transport, although, this context had a much stronger negative impact on a user's ability to assess relevance (as indicated by a significant reduction in number of bookmarks), even more so that observed in a lab setting by \citet{harvey:2018}. The differences in engagement level under these situational contexts indicates that current search features and mobile interfaces do not facilitate search on the go. In spite of the fact that mobile devices are meant to be accessible in almost all situations, users still appear to prefer searching while they are at rest. This suggests that a more dynamic and smart search UI could improve user experience as they walk or are on transport. For example, more voice-based interaction could be adopted, more robust ways of input such as gesture-based input~\cite{DBLP:conf/chi/BragdonNLH11} could be used or the interface could be made to dynamically adapt to context~\cite{harvey:2017}.

%% file: sec-conclusion.tex
\section{Conclusions and Future Work}
\label{sec-conclusion}

The main goal of this study was to understand how temporal and situational contexts affect user behaviour and search performance on mobile devices when performing Web search tasks. We carried out a task-based field study consisting of 31 participants and 12 search tasks in a period of 7 days. We developed \appname, a bespoke Android app, for this study, enabling us to capture various aspects of users' context and behaviour. The participants were given a maximum of two tasks per day at predefined time slots. Randomly distributing the tasks, we were able to cover all different time slots, for different search tasks, on different days. This enabled us to conduct a thorough analysis on the impact of temporal and situational contexts on user behaviour and performance.

Our results indicate that more dynamic interfaces should be designed to encourage users to search while walking or on transport. Also, we observed substantial variation in user behaviour and performance at various times, suggesting that the UI must be aware of the temporal context of the user. Moreover, we saw that users are more willing to complete new search tasks - which presumably require more attention - during early hours of the day. This suggests the need for a more proactive search interface that could provide information cards or suggested queries based on users' regular querying patterns. 

As future work we plan to conduct a follow-up study, focusing on user behaviour as they search in various apps. Recent research has shown that users submit most of their queries within multiple apps~\cite{AliannejadiSigir18,AliannejadiCikm18}. Therefore, it would be interesting to see how users interact with the search engine designed in different apps, under various temporal and situational contexts. Also, as discussed in other studies, user behaviour in terms of app usage vary before, during, and after a search session~\cite{carrascal:2015}. It would also be interesting to study the effect app usage before a search session has on the results that users scan or bookmark. For example, if a user spends a considerable amount of time on YouTube, is he/she more likely to interact with results from \url{http://www.youtube.com} or not?

\begin{acks}
    This research was partially funded by the RelMobIR project of the \grantsponsor{}{Swiss National Science Foundation (SNSF)}{http://www.snf.ch/en/Pages/default.aspx}. We would like to thank the participants who patiently helped us in this study.
\end{acks}